\documentclass{emulateapj}
\usepackage{color}
\usepackage{amsmath}
\usepackage{mathrsfs}
\usepackage{multirow}

\def\be{\begin{equation}}
\def\ee{\end{equation}}
\def\bd{\begin{displaymath}}
\def\ed{\end{displaymath}}
\def\ba{\begin{aligned}}
\def\ea{\end{aligned}}
\def\nms{\mathsurround=0pt}
\def\oversim#1#2{\lower 4pt\vbox{\baselineskip 0pt \lineskip 1pt
    \ialign{$\nms#1\hfil##\hfil$\crcr#2\crcr\sim\crcr}}}
\def\ga{\mathrel{\mathpalette\oversim>}}
\def\la{\mathrel{\mathpalette\oversim<}}

\def\bh{M_{\bullet}}
\def\pc{{\rm \,pc}}
\def\msun{M_{\odot}}
\def\AU{{\rm \,AU}}
\def\kms{{\rm km\,s^{-1}}}

\begin{document}

\title{On the Existence of Pulsars in the Vicinity of the Massive
Black Hole in the Galactic Center}
\shortauthors{Zhang, Lu \& Yu}

\author{Fupeng Zhang$^{1,2}$, Youjun Lu$^1$ \& Qingjuan Yu$^{2}$}
\affil{$^1$~National Astronomical Observatories, Chinese Academy of
Sciences, Beijing, 100012, China; luyj@nao.cas.cn \\
$^2$~Kavli Institute for Astronomy and Astrophysics, Peking
University, Beijing, 100871, China; zhangfupeng, yuqj@pku.edu.cn
}

\begin{abstract}

Pulsars, if existing and detectable in the immediate vicinity of the
massive black hole (MBH) in the Galactic center (GC), may be used as a
superb tool to probe both the environment and the metric of the central
MBH. The recent discovery of a magnetized pulsar in the GC suggests
that many more pulsars should exist near the MBH.  In this paper, we
estimate the number and the orbital distribution of pulsars in the
vicinity of the MBH in the GC by assuming that the pulsar progenitors,
similar to the GC S-stars, were captured to orbits tightly bound to
the MBH through the tidal breakup of stellar binaries. We use the
current observations on both the GC S-stars and the hypervelocity
stars to calibrate the injection rate(s) of and the dynamical model(s)
for the stellar binaries.  By including the relaxation processes,
supernova kicks, and gravitational wave radiation in our simulations,
we estimate that $\sim97-190$ ($9-14$) pulsars may presently orbit
the central MBH with semimajor axes $\leq4000\AU$ ($\leq1000\AU$),
which is compatible with the current observational constraints on the
number of the GC pulsars. The semimajor axis and the pericenter
distance of the pulsar closest to the central MBH are probably in the
range of $\sim120-460\AU$ and $\sim2-230\AU$, respectively. Future
telescopes, such as the SKA, may be able to detect a significant
number of pulsars with semimajor axis smaller than a few thousand AU
in the GC. Long-term monitoring of these pulsars would be helpful in
constraining both the environment and the metric of the central MBH.
Our preferred model also results in about ten hyperfast pulsars with
velocity $\ga1500\kms$ moving away from the Milky Way.

\end{abstract}


\section{Introduction}

Long term monitoring of the motion of the S-stars\footnote{Hereafter
denoted as the GC S-stars.} located at the Galactic center (GC)
demonstrates exclusively the existence of a massive black hole (MBH)
in the GC \citep[e.g.,][]{Ghezetal08, Gillessenetal09}. Although the
source confusion in observations has so far inhibited the discovery of
any star rotating around the MBH with an orbital period $<10$~yr
\citep{Meyer12}, young stars and compact stars, including neutron
stars and stellar mass black holes, are expected to exist in this
region \citep[e.g.,][]{MG00, Angelil10, liu12, ZLY13}. If some of
these stars are radio pulsars, which would be promising in
observational detection and providing a superb clean tool to probe the
gravitational field of the central MBH \citep[e.g.,][]{PT79, Pfahl04,
Angelil10}. It becomes especially encouraging that a magnetar, a
highly magnetized pulsar, is recently discovered at a parsec (pc)
distance from the central MBH. This magnetar has been used to reveal
the magnetic field in the GC \citep[see][]{Rea13, Eatough13, Kennea13,
Mori13}. Magnetars are a rare type of pulsars, and the discovery of a
magnetar in the GC suggests that many more normal pulsars exist there.

The GC S-stars may be the captured components of stellar binaries that
were tidally broken up in the vicinity of the central MBH as proposed
by \citet[][see also \citealt{GL06, Lockmann08, Perets09, AM13,
ZLY13}]{Gould03}.  The other components of those broken-up binaries
may be ejected to the Galactic halo as hypervelocity stars
\citep[HVSs; e.g.,][]{Hills88, YT03, Brown05, Edelmann05, Hirsch05,
Brown12}.  \citet[][hereafter ZLY13; see also \citealt{LZY10}, and
\citealt{ZLY10}]{ZLY13} find that the current observations on both the
GC S-stars and the HVSs are compatible with the model that their
progenitorial stellar binaries are originated from the young stellar
disk(s) at a distance of $0.04-0.5\pc$ from the MBH
\citep{LuJ09,Bartko09}.

Assuming that both the HVSs and the GC S-stars are produced by the
tidal breakup mechanism, the injection rate of stellar binaries to the
immediate vicinity of the central MBH can be constrained by using
current observations on the numbers and distributions of these two
populations.  Some of the captured/ejected massive components (with
mass $\geq 9\msun$; \citealt{Heger03}) of the broken-up binaries may
collapse and explode to form pulsars at the end of their main-sequence
lives. Therefore, the number of the pulsars currently existing in the
GC can be reasonably estimated if adopting the tidal breakup mechanism
to produce the massive GC S-stars.  \citet{Pfahl04} estimated the
number of the pulsars existing in the vicinity of the central MBH
($\la 4000\AU$)\footnote{Different units for the distance of a star
from the GC MBH are chosen in different references, e.g., pc in
\citet{liu12} and \citet{Gillessenetal09}, mas ($10^{-3}\,$arcsec) in
\citet{Ghezetal08}, and gravitational radius $r_{\rm g}$ in
\citet{Angelil10}, respectively. For the GC MBH system, we note here
that $1 \text{mas} \simeq 3.88\times 10^{-5}\,\text{pc} (R_{\rm GC}/8
\text{kpc}) \simeq 8\text{AU} (R_{\rm GC}/8 \text{kpc}) \simeq 202
r_{\rm g} (M_{\bullet}/4\times 10^6\msun)^{-1} (R_{\rm GC}/8\text{kpc})$.
In this paper, the MBH mass is set to be $M_{\bullet} = 4\times
10^6\msun$, the distance from the GC to the sun is $R_{\rm GC} = 8
\text{kpc}$, and we mainly use the physical distance unit AU to
describe the distance of a GC star from the central MBH.} to be $\sim
1000$ (see also a new estimate on the pulsar generation rate in the GC
by \citealt{DO13}). In this paper, we revisit the estimation by
including more detailed and realistic considerations of various
physical processes involved in, such as the dynamical evolution of the
pulsars and their progenitors, the kicks received by the pulsars at
their birth, and the orbital decay due to the gravitational wave (GW)
radiation, etc.

This paper is organized as follows. In Section~\ref{sec:simu}, we
perform a large number of full three-body experiments to realize the
tidal breakup processes of stellar binaries in the vicinity of the
central MBH by adopting different binary injection models. We follow
the dynamical evolution of the captured components, by considering the
resonant and the non-resonant relaxations due to background stars, and
the decay due to the GW radiation.  If a captured (or an ejected) star
ends its main-sequence life and explodes to form a pulsar before the
end of the simulation, the motion of the pulsar is assumed to receive
a kick at its birth, based on a kick-velocity distribution obtained
from observations. We set the end of the simulations as the present
time. We follow the kinematic motions of the ejected components in the
Galactic potential and obtain the number and the  spatial distribution
of HVSs at the present time. The parameters describing the initial
settings for each hypothesized injection model can be calibrated by
using current observations as done in ZLY13. With the calibrated model
parameters, the number and the spatial distribution of these pulsars
are estimated in Section~\ref{sec:result}. We further estimate the
probability distributions of the semimajor axis and the pericenter
distance of the inner most pulsar. Discussion and conclusions are
given in Section~\ref{sec:conclusion}.

\section{Monte-Carlo Simulations}\label{sec:simu}

\begin{deluxetable*}{lccclllclll}
\tablehead { \multirow{2}{*}{Model} & \multirow{2}{*}{$\gamma$} & \multirow{2}{*}{$\beta$} & \colhead{} &
\multicolumn{3}{c}{core-like profile ($\alpha=0.5$)} & \colhead{} &
\multicolumn{3}{c}{Bahcall-Wolf cusp ($\alpha=7/4$)} \\
\cline{5-7} \cline{9-11} 
   &   &  &  &
\colhead{$N^{\rm obs}_{\rm P}$} & \colhead{$N^{\rm obs}_{\rm ej}$} & \colhead{$N^{\rm obs}_{\rm kick}$}  & 
& \colhead{$N^{\rm obs}_{\rm P}$} & \colhead{$N^{\rm obs}_{\rm ej}$} & \colhead{$N^{\rm obs}_{\rm kick}$} 
}
\startdata
Unbd-MS0   &-2.70 & 0  & & 38\ \ \ (52)  & 33 (47)     & 0 (0) & & 31\ \ \ (43)   & 45 (59)     & 0 (0)\\
Disk-TH0   &-0.45 & 0  & & 96\ \ \ (391) & 54 (199)    & 1 (2) & & 76\ \ \ (336)  & 62 (241)    & 1 (2) \\
Disk-TH2   &-0.45 & 2  & & 128    (470)  & 8\ \ \ (31) & 2 (7) & & 100     (406)  & 8\ \ \  (36)& 3 (6) \\
Disk-IM0   &-1.60 & 0  & & 61\ \ \ (126) & 33 (61)     & 1 (1) & & 48\ \ \ (100)  & 36 (69)     &  0 (0)\\
Disk-IM2   &-1.60 & 2  & & 97\ \ \ (190) & 4\ \ \ (8)  & 2 (3) & & 69\ \ \ (152)  & 5\ \ \ (9)  & 3 (4) 
\enddata
\label{tab:t1}
\tablecomments{
Numbers of the ejected and the captured pulsars at the present time
obtained from different models.  The injection rate of stellar
binaries is assumed to be a constant over the past 250 Myr, which is
calibrated by producing $17$ GC S-stars with semimajor axis $<4000\AU$
for each model.  $N^{\rm obs}_{\rm P}$, $N^{\rm obs}_{\rm ej}$ and
$N^{\rm obs}_{\rm kick}$ represent the number of pulsars with
semimajor axis $\leq 4000\AU$ at the end of the simulation, the total
number of pulsars with velocity $\geq 1500\kms$ at infinity, and the
number of those pulsars with velocity $\geq 1500\kms$ at infinity that
are originally simulated GC S-stars but were kicked out from the GC
due to supernova explosions, respectively. The numbers listed in the
fourth column to sixth column and the seventh column to ninth column 
are obtained by adopting a core-like density
profile and a Bahcall-Wolf cusp for the background stars,
respectively.  From the fourth to ninth columns, the numbers within (or
out of) brackets represent the results obtained by assuming that those
simulated GC S-stars with mass $\geq 9\msun$ (or in the range of
$9\msun-25\msun$) can turn to pulsars when they end their main-sequence life.  }
\end{deluxetable*}

\subsection{Model Settings}
\label{subsec:modelset}

We first perform a large number of Monte-Carlo three-body experiments
to realize the tidal breakup processes of stellar binaries in the
vicinity of the central MBH and generate the mock samples of the GC
S-stars and the HVSs. We adopt the code DORPI5 based on the explicit
fifth(fourth)-order Runge-Kutta method to simulate the three-body
interaction between a stellar binary and the MBH \citep{DP80,Haier93}.
For details of the simulation, see \citet{ZLY10}.
Similar to ZLY13, we also adopt
several injection models with different settings on the origin of the
stellar binaries and the distributions of their properties, denoted as
``Unbd-MS0'', ``Disk-TH0'', ``Disk-TH2'', ``Disk-IM0'', and
``Disk-IM2'', respectively (see Table~\ref{tab:t1}, and also Section
3.1 in ZLY13). 

\begin{itemize}

\item The progenitorial stellar binaries are assumed to be
originated either from disk structure(s) like the clockwise rotating
stellar disk in the GC (the last four models) or from infinity with
initial velocities of $250\kms$ (the first model), which are
represented by the ``Disk'' and ``Unbd'' in the model notations,
respectively. 

\item
Current observations show that the initial mass function (IMF) of the
young stars in the GC is $-1.7$ \citep[see][]{LuJ13} or $-0.45$
\citep[see][]{Bartko10}.  The IMF estimated by \citet{LuJ13} is
consistent with the constraint obtained from the number ratio of the
HVSs to the GC S-stars if both of them are produced by the tidal
breakup of stellar binaries originated from the stellar disk(s) in the
GC (see ZLY13). We adopt the Miller-Scalo IMF ($\gamma=-2.7$) for the
Unbd-MS0 model, an IMF with a slope of $\gamma=-1.6$ (denoted as an
intermediate slope) for the Disk-IM0 model and the Disk-IM2 model, and
a top-heavy IMF with $\gamma=-0.45$ for the Disk-TH0 model and the
Disk-TH2 model, which are denoted as the ``MS'', ``IM'' and ``TH'' in
the model notations, respectively. The number ``0'' and ``2'' in the
model notations represent that the distribution of the initial
pericenter distances $r_{\rm p,i}$ of the injecting binaries
approaching the MBH follows a power law, i.e., $\propto
r_{\rm p,i}^{\beta}$, with a slope $\beta=0$ or $\beta=2$,
respectively. The total number of three-body experiments is $10^5$ for
each model.  For details of the initial settings and other properties
of the stellar binaries in these injection models, see ZLY13.

\end{itemize}

We assume that the stellar binaries injected (or migrated) into the
vicinity of the central MBH and were tidally broken-up soon after
their formation, therefore, all stars were on the main-sequence before
they were captured by the MBH or ejected from the GC.  We assume a
constant injection rate of stellar binaries over the past $250$Myr in
each model, which can be calibrated by the observational statistical
properties of the GC S-stars (and/or the HVSs) as done in ZLY13. We
follow the dynamical evolution of the captured components by adopting
the Auto-Regressive Moving Average (ARMA) model given by
\citet{MHL11}, in which both the resonant and the non-resonant
relaxations are considered (see also Section~4 in ZLY13 for details).
The background stars are the main objects that cause both the resonant
and the non-resonant relaxations of the captured GC S-stars and
pulsars.  Recent observations suggest that the density of those
background stars can be modeled by a core density profile
\citep[e.g.,][]{Doetal09}.  Therefore, we adopt a power-law density
distribution for the background stars, i.e., $\rho_*\propto
r^{-\alpha}$, with $\alpha=0.5$, to mimic the effect of a core-like
distribution (see also \citealt{MHL11} and ZLY13). We will also
discuss the effect on the results by choosing a Bahcall-Wolf cusp
$\alpha=7/4$ \citep{BW76} in Section~\ref{sec:result}. 

In our calculations, we remove those captured stars that approach the
MBH within the tidal radius, i.e., $(2\bh/m_*)^{1/3} R_*$, where $m_*$
and $R_*$ are the stellar mass and radius, because they should be
tidally disrupted and swallowed by the central MBH.  We also follow
the kinetic motions and evolution of the ejected components in the
Galactic potential (for details see Section 5 in ZLY13).

\subsection{GC pulsars and Supernova Kick}
\label{sebsec:SNkick}

The captured massive stars may evolve into pulsars/neutron stars after
their main-sequence lives. According to \citet{Heger03} and
\citet{Georgy09}, stars with mass in the range of $9-25\msun$ can turn
into pulsars/neutron stars, while some with mass above $25\msun$ stars
may turn into pulsars/neutron stars only if their metallicities are
above the solar abundance. Observational constraints on the progenitor
mass of pulsars are more or less consistent with theoretical
expectations \citep[e.g.,][]{Smartt09}.  It has been shown that the
metallicities of the GC S-stars may be higher than the solar abundance
\citep[e.g.,][]{Paumard06}. Therefore, we consider two extreme cases
in this study: (a) only those simulated GC S-stars and HVSs with mass
in the range $9-25\msun$ can evolve into pulsars once they end their
main-sequence lives; and (b) all those simulated GC S-stars and HVSs
with mass $\geq 9\msun$ can turn into pulsars once they end their
main-sequence lives. We assume the active time of these pulsars is
$\sim 100$~Myr.  In this study, we mainly focus on case (a) if not
otherwise stated; the numbers of pulsars resulting from case (a) are
taken as the reference numbers, and those from case (b) may be taken
as the upper limits. 

A pulsar formed through the collapse of a massive star may receive a
kick at its birth. According to pulsar observations, the
one-dimensional component of this kick velocity follows an exponential
distribution as estimated by \citet{FK06}, i.e.,
\be
P(v_l)=\frac{1}{2\left<|v_l|\right>} \exp\left(-\frac{|v_l|}{\left<|v_l
|\right>}\right), 
\label{eq:vkick}
\ee
where $v_l$ is the one-dimensional component of the 3D velocity,
$\left<|v_l|\right>=180^{+20}_{-30} \kms$.\footnote{Adopting a different
kick velocity distribution [i.e., as that estimated by
\citet{Arzoumanian02}, or \citet{CC98}, or \citet{LL94}, or
\citet{HP97}], the resulting numbers of active pulsars may increase or
decrease by only $\la 20\%$.} According to Equation~(\ref{eq:vkick}),
we randomly assign each component of a 3D kick velocity $\delta
\vec{v}$ to each newly born pulsar, and consequently the directions of
the 3D kick velocities for pulsars are randomly distributed. By
receiving such a kick, the specific energy $E$ and angular momentum
$\vec{J}$ of a pulsar change by
\be
\delta E_{\rm kick}=\vec{v}_{\rm orb}\cdot \delta \vec{v} + 
\frac{1}{2} |\delta \vec{v}|^2,
\label{eq:energy}
\ee
and
\be
\delta \vec{J}_{\rm kick}=\vec{r}_{\rm orb}\times \delta\vec{v},
\label{eq:ang}
\ee
respectively, where $\vec{r}_{\rm orb}$ and $\vec{v} _{\rm orb}$ are
the position vector and the orbital velocity of the pulsar,
respectively. The changes of the
specific energy and angular momenta due to the supernova kick are
included in the ARMA model according to 
Equations~(\ref{eq:energy}) and (\ref{eq:ang}) above.

\subsection{Orbital Decay due to the Gravitational Wave Radiation}
\label{sebsec:GWdecay}

For those pulsars with extremely high orbital eccentricities $e$ and
small semimajor axes $a$, the GW radiation may be important for their
orbital evolution. The GW decay timescale for an object, with mass
$m$, eccentricity $e$, and semimajor axis $a$, is given by
\citep{Peters64}
\be
\ba
T_{\rm GW}  &\sim 1.2\times10^{8}{\rm yr}\  \frac{f(0.99)}{f(e)}
                      \left(\frac{1.4\msun}{m}\right) \\
            &\times \left(\frac{a}{300\AU}\right)^4 \left(\frac{\bh}{4
             \times 10^6\msun}\right)^{-3},
\ea
\ee
and
\be
f(e)\equiv (1-e^2)^{-7/2}\left(1+\frac{73}{24}{e^2}
+\frac{37}{96}{e^4}\right).
\ee
The GW decay timescale $T_{\rm GW}$ can be comparable to the active
time of pulsars if $m\sim 1.4\msun$, $a\sim 300\AU$ and $e \sim 0.99$.
We modify the ARMA model by adding additional terms in the evolution
of their orbital energy $E$ and momenta $J$ in order to include this 
GW effect on the orbital evolution of the GC pulsars (and also the
GC S-stars) \citep[see][]{Peters64}, i.e.,
\be
\delta E=P\frac{dE}{dt} =-\frac{32}{5}\frac{G^4m\bh^3}{c^5 a^5}
\frac{1+\frac{73}{24}e^2+\frac{37}{96}e^4}{(1-e^2)^{7/2}}
\times P,
\ee
\be
\delta J= P \frac{dJ}{dt} =-\frac{32}{5}
\frac{G^{7/2}m\bh^{5/2}}{c^5 a^{7/2}}
\frac{1+\frac{7}{8}e^2}{(1-e^2)^{2}}\times P,
\label{eq:6_GW}
\ee
where $G$ is the gravitational constant, $c$ is the light speed, and
$P=2\pi (a^3/G \bh)^{1/2}$ is the orbital period of a pulsar (or a GC
S-star). 

For the secular evolution of the orbits of pulsars/stars, we consider
the periastron advances due to the general relativity correction and
the orbital decay due to the gravitational radiation in the ARMA
model. We neglect the effects due to other low-order post-Newtonian
terms (such as the Lense-Thirring precession and the frame dragging,
etc.) for the following reasons.  (1) The magnitude of the
Lense-Thirring precession is relatively much smaller than that of the
periastron advance, and the periastron advance has been included in
the ARMA model (see ZLY13 and \citealt{MHL11}); (2) the other
low-order post-Newtonian effects do not lead to secular orbital decay
of those pulsars; and (3) although those effects may affect the
orbital motion of individual stars in the vicinity of the MBH
\citep[e.g., see][]{Iorio11,Angelil10}, we obtain the evolution of the
GC system in a {\it statistical} way through the ARMA model, which
should be sufficient for the purpose of the study presented here. 

\begin{figure*}
\center
\includegraphics[scale=0.85]{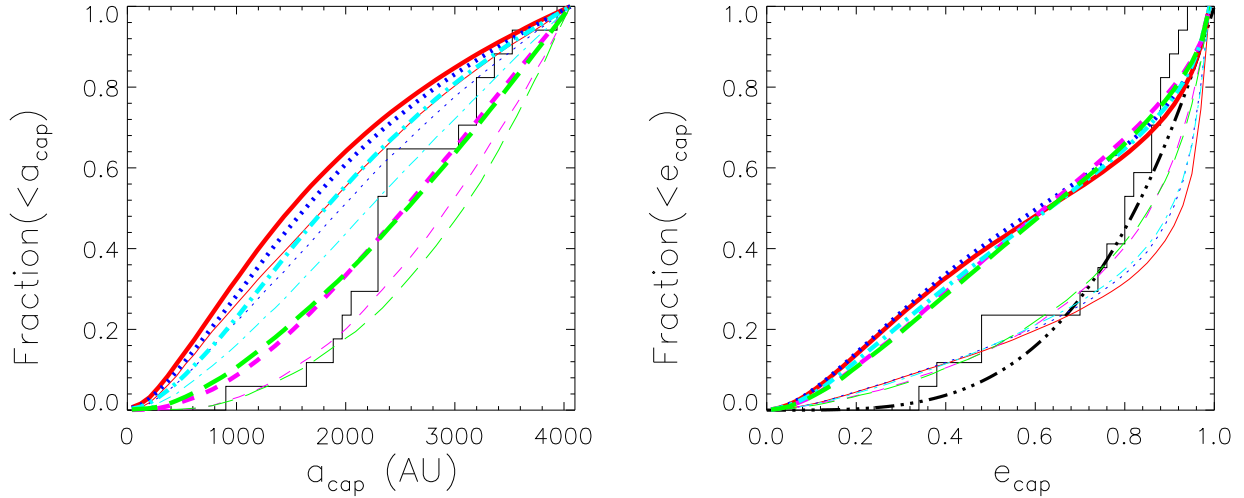}

\caption{Cumulative distributions of the semimajor axis (left panel)
and the eccentricity (right panel) of the captured stars (thin lines)
and pulsars (thick lines) with semimajor axis $\leq 4000$~AU surviving
to the present time, respectively. The solid (red), dotted (blue),
dashed (magenta), dot-dashed (cyan) and long-dashed (green) curves
represent the results obtained from the Unbd0-MS model, the Disk-TH0
model, the Disk-TH2 model, the Disk-IM0 model, and the Disk-IM2 model,
respectively. In each panel, the histogram represents the distribution
of the observed GC S-stars \citep{Gillessenetal09}.  In the right
panel, the black dot-dot-dot-dashed line represents a cumulative
distribution proportional to $e_{\rm cap}^{3.6}$ as suggested by the
observations \citep{Ghezetal08, Gillessenetal09}.  }

\label{fig:f1}
\end{figure*}

We trace the motions of those newly born pulsars until the end of our
simulations. For each pulsar, the calculation is terminated if its age
reaches $100$~Myr (which is assumed to be the lifetime of active
pulsars) or if it is swallowed by the central MBH.  If the lifetime of
active pulsars is shorter than the assumed one, the number of the
resulting pulsars decreases proportionally. A pulsar can directly
plunge into and be swallowed by the central MBH if its angular
momentum is less than $4G\bh/c$ (and thus on an unstable orbit; e.g.,
\citealt{Chandrasekhar83}, \citealt{Ivanov02}), although the tidal
radius of the pulsar is smaller than the MBH event horizon. We remove
those pulsars with angular momenta $\leq 4G\bh/c$ from our
calculations.

\section{Results}\label{sec:result}

We record the simulated GC S-stars, HVSs and active pulsars at the end
of the Monte-Carlo simulations for each model.  Figure~\ref{fig:f1}
shows the cumulative distributions of the semimajor axis and
eccentricity for both the simulated GC S-stars (thin lines) and the
pulsars (thick lines) obtained from different models. As seen from
Figure~\ref{fig:f1}, the resulting distributions of the semimajor axis
and eccentricity of the simulated GC S-stars are more or less
statistically compatible with the observations (see also discussions
in ZLY13). Further in consideration of the ratio of the number of the
HVSs to that of the GC S-stars and the velocity distribution of HVSs,
the Disk-IM2 model may match the observations better (see also
detailed analysis in ZLY13). The distributions of the semimajor axis
and the eccentricity of the active pulsars appear to be quite
different from those of the GC S-stars, which is mainly due to: (1)
the supernova kicks received by pulsars at their birth, and (2) the
dynamical evolution timescale of the active pulsars ($10^8$~yr) is
substantially longer than those of the GC S-stars (at most a few times
$10^7$~yr).

For each model, we count the total number of those simulated GC
S-stars with mass in the range of $\sim 7-15\msun$ and semimajor axes
$a \leq 4000$\AU, denoted as $N^{\rm sim}_{\rm S}$; and we also count
the final number of those simulated active pulsars with $a \leq
4000$\AU, denoted as $N^{\rm sim}_{\rm P}$.  As the total number of
the detected GC S-stars with $a \leq 4000$\AU\ is known to be $N^{\rm
obs}_{\rm S}=17$ \citep{Gillessenetal09}, therefore, the total number
of active pulsars at the present time can be estimated as $N^{\rm
obs}_{\rm P} = N^{\rm obs}_{\rm S} N^{\rm sim}_{\rm P} /N^{\rm
sim}_{\rm S}$ for each model.  By doing so, we estimate the total
number of active pulsars (with $a \leq 4000$ \AU) at the present time
for each model as listed in Table~\ref{tab:t1}.  The Disk-IM2 model
may match the statistical properties of both the GC S-stars and the
HVSs better as demonstrated in ZLY13, therefore, the total number of
active pulsars with semimajor axes $\leq 4000\AU$\ (or $\leq 1000\AU$)
is $\sim 97-190$ (or $\sim 9-14$) in the preferred DISK-IM2 model if
choosing a core-like density profile for the background stars. If
choosing a Bahcall-Wolf cusp for the background stars, this number
decreases slightly to $\sim 69-152$ (or $\sim 4-7$). The decrease of
the number of the resulting pulsars is due to that the resonant
relaxation process is more efficient in the case of a Bahcall-Wolf
cusp than that of a core-like density profile, which leads to more
pulsars being excited to extremely high eccentric orbits and being
swallowed by the central MBH (see also discussions in ZLY13).  The
estimated number is smaller than the previous estimation ($\sim 1000$
by \citealt{Pfahl04}) by one order of magnitude. The main reason for
this difference is that we further consider the dynamical model of
capturing stars by the central MBH in details, the effect of the IMF,
the dynamical evolution of the GC S-stars and pulsars, and the effect
of the kick received by a newly formed pulsar at its birth, etc.  For
example, more than $25\%$ of the pulsars are ejected out from the
immediate vicinity of the central MBH to infinity because of the 
kicks received by these pulsars at their birth.

\begin{figure*}
\center
\includegraphics[scale=0.7]{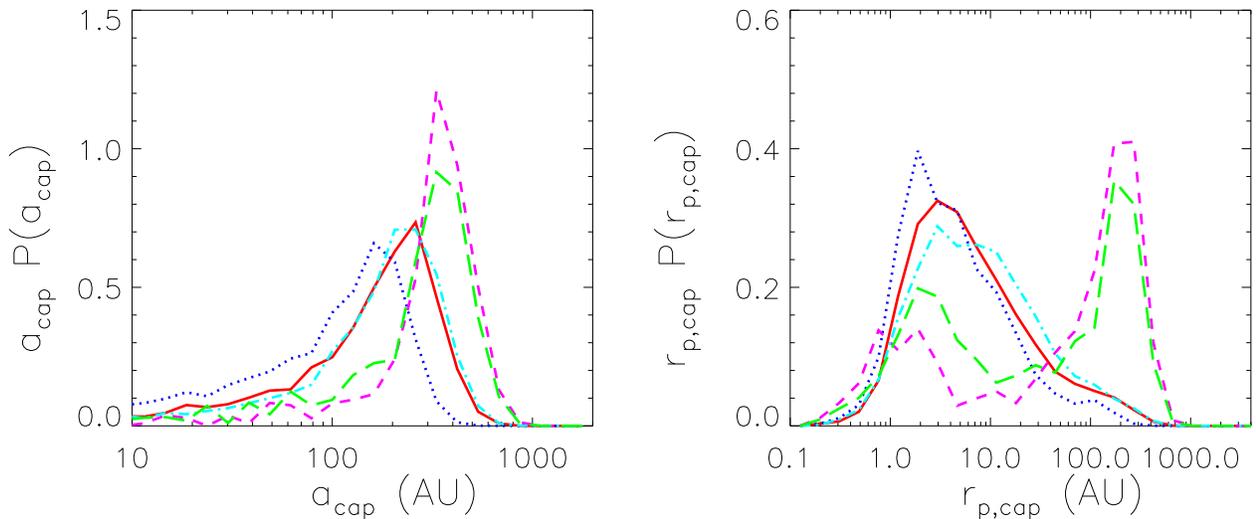}

\caption{Probability density distributions of the semimajor axis (left
panel) and the pericenter distance (right panel) of the innermost
pulsar at the present time. The solid (red), dotted (blue), dashed
(magenta), dot-dashed (cyan) and long dashed (green) curves represent
the results obtained from the Unbd-MS0 model, the Disk-TH0 model, the
Disk-TH2 model, the Disk-IM0 model, and the Disk-IM2 model,
respectively. The estimates shown here are obtained by (1) assuming
that the captured stars with mass in the range of $9 - 25 \msun$ all
turn into pulsars after their main-sequence lives; and (2) calibrating
the injection rate of stellar binaries over the past $250$~Myr to
generate 17 GC S-stars at the present time, the same number as that of
the observed ones.  }

\label{fig:f2}
\end{figure*}

\subsection{The Innermost Pulsar}
\label{subsec:innerpulsar}

The pulsar with the smallest semimajor axis (and pericenter distance),
if detectable, should be most helpful in probing the strong
gravitational field and constraining the metric of the central MBH. In
order to check whether it is possible to find a pulsar that is
sufficiently close to the central MBH, we estimate the probability
distributions of the semimajor axis and the eccentricity of the
innermost pulsar resulting from each model above. To do so, we first
randomly select $N^{\rm obs}_{\rm P}$ pulsars from those active
pulsars resulting from each model and record the properties of the one
with the smallest semimajor axis among them. We repeat this procedure
by $10^6$ times, and then use the recorded innermost pulsars to obtain
the probability distributions of the semimajor axis and the pericenter
distance for the innermost active pulsar. 

Figure~\ref{fig:f2} shows the probability density distributions of the
semimajor axis (left panel) and the pericenter distance (right panel)
of the innermost pulsar resulting from different models, where $P(x)$
is defined so that $P(x) dx$ is the probability of being in the range
of $x\rightarrow x+dx$.  As seen from Figure~\ref{fig:f2}, the
Disk-TH0 model results in an innermost pulsar relatively closer to the
central MBH than all the other models.  The semimajor axis
distributions obtained from the Unbd-MS0 model and the Disk-IM0 model
(or from the Disk-TH2 model and the Disk-IM2 model) are similar; and
the models with small $\beta$ ($=0$) result in an innermost pulsar
more closer to the central MBH than the models with large $\beta$
($=2$; see the left panel of Figure~\ref{fig:f2}), as more binaries
were injected into the immediate vicinity of the central MBH in the
models with $\beta=0$ than that in the models with $\beta=2$.

Note there are two peaks in the pericenter distance distribution of
the innermost pulsar resulting from the Disk-TH2 model and the Disk-IM2
model (see the short dashed and long dashed lines in the right panel of
Figure~\ref{fig:f2}).  The left peak (close to the MBH) is due to that
the orbits of a number of pulsars, with small semimajor axes and large
eccentricities (e.g., $a_{\rm cap} \sim 300\AU$ and $e \ga
0.99$), decay rapidly due to the GW radiation, while the right peak
(further away from the MBH) represents those pulsars that are not
significantly affected by the GW radiation.  

As a comparison to the dynamical evolution of the GC pulsars, the GW
radiation plays little role in the orbital evolution of the simulated       
GC S-stars. The reasons for this are two folds: first, the lifetime of      
the simulated GC S-stars is only $\sim 10^7$~yr, which is much less
than the GW decay timescale as these stars typically have semimajor
axes larger than a few hundred \AU; second, for those simulated GC
S-stars that can be excited to extremely highly-eccentric orbits ($e >        
0.99$), their orbits decay fast due to the GW radiation but they are        
more likely to be tidally disrupted and swallowed by the central MBH        
rather than survive to the present time. 

We define the most probable range of the semimajor axis (or the
pericenter distance) of the innermost pulsar here as the range from
$16$ percentile to $84$ percentile of the distributions, which is $57
- 303\AU$ ($2 - 24\AU$), $31 - 211\AU$ ($1 - 14\AU$), $199 - 461\AU$
($2 - 256\AU$), $76 - 322\AU$ ($2-30\AU$), and  $116 - 461\AU$ ($2 -
  228\AU$), for the Unbd-MS0 model, the Disk-TH0 model, the Disk-TH2
model, the Disk-IM0 model, and the Disk-IM2 model, respectively. For
all the models, including the reference model (Disk-IM2), the
semimajor axis of the resulting innermost pulsar is $\la 300 - 500\AU$
(corresponding to a period of $\sim 2.6-5.6$~yr), substantially
smaller than that of the currently known GC S-stars. According to the
probability distributions shown in Figure~\ref{fig:f2}, the
probability of the innermost pulsar with semimajor axis $\leq 300\AU$
(or $\leq 100\AU$) is $83\%$ ($26\%$), $98\%$ ($43\%$), $33\%$
($8\%$), $79\%$ ($22\%$), and $45\%$ ($15\%$), for the Unbd-MS0 model,
the Disk-TH0 model, the Disk-TH2 model, the Disk-IM0 model, and the
Disk-IM2 model, respectively. The existence of such a pulsar, if
detectable, should provide a superb tool to probe the metric and
determine the spin of the central MBH. As pointed out by \citet[][see
also \citealt{Angelil10, Pfahl04}]{liu12} that the frame-dragging
signal or even higher order general relativistic effects could be
detected by using the pulsar timing of a pulsar orbiting the central
MBH with a semimajor axis smaller than a few hundred AU if the timing
accuracy is moderately high (e.g., $\la 10^{-4}$\,s).

\begin{figure*}
\center
\includegraphics [scale=0.7]{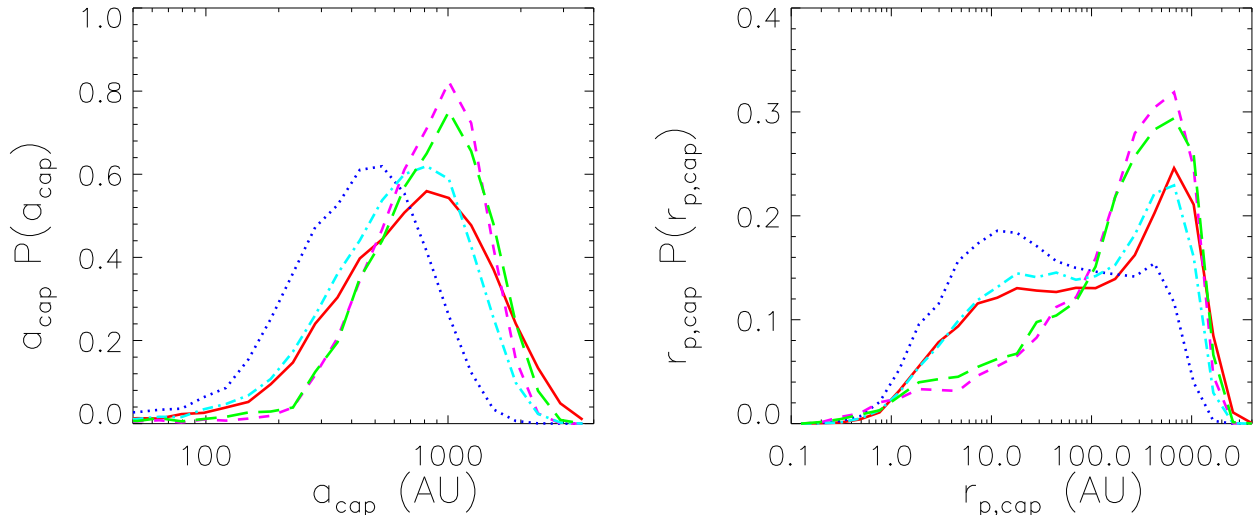}

\caption{Legend similar to Figure~\ref{fig:f2}, except that it is for
the innermost ``detected'' pulsar by assuming that only $10\%$ of the
simulated pulsars can be detected.  }

\label{fig:f3}
\end{figure*}

In fact, not every pulsar can be detected because of many factors,
including the beaming effect and the scattering of radio wave by
turbulent ionized gas between the pulsar and the observer. Assuming
that optimistically $10\%$ of the normal pulsars can be detected by
future facilities, such as the Square Kilometer Array (SKA) \citep[see
Equation~2 in][]{Pfahl04}, we also estimate the probability
distributions of the semimajor axis and the eccentricity of the
innermost ``{\it detected}'' pulsar for each model, which are shown in
Figure~\ref{fig:f3}.  The most probable range of the semimajor axis
(pericenter distance) of the innermost ``{\it detected}'' pulsar is
$340 - 1424\AU$ ($7 - 791\AU$), $211 - 754\AU$ ($4 - 286\AU$), $468 -
1350\AU$ ($27 - 791\AU$), $306 - 1151\AU$ ($8 - 631\AU$), and $468
- 1423\AU$ ($21 - 791\AU$) for the Unbd-MS0 model, the Disk-TH0 model,
  the Disk-TH2 model, the Disk-IM0 model, and the Disk-IM2 model,
respectively. For majority of the models, the semimajor axis of the
resulting innermost pulsar is probably still smaller than that of the
currently known GC S-stars. However, the probability of the
``detected'' innermost pulsar with semimajor axis $\leq 300\AU$ (or
$\leq 100\AU$) is only $13\%$ ($2\%$), $31\%$ ($4\%$), $4\%$
($0.6\%$), $15\%$ ($2\%$), and $5\%$ ($1\%$) for the Unbd-MS0 model,
the Disk-TH0 model, the Disk-TH2 model, the Disk-IM0 model, and the
Disk-IM2 model, respectively. As the probability is not negligible,
there might be still some chance to find a pulsar sufficiently close
to the central MBH by future facilities that may help to constrain the
MBH metric.

For case (b), i.e., all captured stars with mass $>9\msun$ turn to
active pulsars when they end their main-sequence life, then the
probability that the innermost ``detected'' pulsar has a semimajor
axis $\leq 300\AU$ is $85\%$ (or $16\%$), $100\%$ ($64\%$), $78\%$
($14\%$), $97\%$ ($27\%$), and $52\%$ ($7\%$) for the Unbd-MS0 model,
the Disk-TH0 model, the Disk-TH2 model, the Disk-IM0 model, and the
Disk-IM2 model, respectively, if all (or only $10\%$ of) the resulting
pulsars can be detected. These probabilities are larger than that
obtained for case (a) above, which may strengthen the above conclusion
that a pulsar substantially close to the central MBH may be found by
future facilities and help to constrain the MBH metric.  

If alternatively adopting a Bahcall-Wolf cusp for the background stars
for case (a), the probability of the innermost ``detected'' pulsar
with semimajor axis $\leq 300\AU$ is $42\%$ (or $5\%$), $74\%$
($12\%$), $18\%$ ($2\%$), $43\%$ ($5\%$), and $34\%$ ($3\%$) for the
Unbd-MS0 model, the Disk-TH0 model, the Disk-TH2 model, the Disk-IM0
model, and the Disk-IM2 model, respectively, if all (or only $10\%$
of) the resulting pulsars can be detected. These resulting
probabilities are only slightly different from that obtained above by
adopting a core-like density profile for the background stars.
Therefore, qualitatively our main results are not affected by choosing
a Bahcall-Wolf cusp for the background stars.

The pulsars generated above are all formed by the collapse and
explosion of the captured components of stellar binaries that were
tidally broken up in the vicinity of the central MBH. It is possible
that other mechanism(s) may also bring pulsars to the immediate
vicinity of the central MBH. For example: (1) some binaries with a
pulsar component (or even two pulsar components) may be injected into
the vicinity of the MBH and then tidally broken up with the pulsar
component being captured. However, most of newly formed pulsars should
be kicked out from binary systems by receiving the kicks due to
supernovae explosion, and stay in isolate far away from the central
MBH. It is reasonable to simply ignore the contribution by this
mechanism. (2) Some isolated pulsars formed in the outer region of the
GC \citep[e.g.,][]{Zubovas13,Baruteau11} may migrate into the
immediate vicinity of the central MBH via two-body relaxation.  The
typical timescales of the two-body relaxation in the S-star cluster
region ($\sim 10^9$~yr or larger; see \citealt{HA06,YLL07}) are far
larger than the active time of pulsars ($\sim 10^8$yr). Therefore, the
active pulsars within $4000\AU$ from the MBH should not be dominated
by the pulsars resulting from the above two ways. In summary, our
results on the number and orbital distribution of the pulsars with
semimajor axis $\leq 4000\AU$ are not significantly affected by
considering the above two mechanisms.

\subsection{Hyperfast Pulsars}
\label{subsec:hyperfastpulsar}

Some hyperfast pulsars can also be produced by the process of tidal
breakup of stellar binaries in the vicinity of the central MBH
\citep[see][]{GGP08}. There are two routes to form such hyperfast
pulsars in the scenario of tidal breakup of stellar binaries.

\begin{itemize}

\item Collapse and explosion of massive HVSs. In our simulations, many
HVSs, with mass $\geq 9\msun$, ended their main-sequence lives and
collapsed to form pulsars before the end of the calculations. The
ejection velocities of those HVSs are already high (e.g., $\ga
1000\kms$). The pulsars formed from those HVSs can be further
accelerated to even high velocities (e.g., $\ga 1500\kms$) by the
kicks received at its birth if the directions of these kicks
are close to the velocity vectors of the progenitorial
HVSs.

\item Collapse and explosion of some massive captured stars in the
vicinity of the central MBH. The pulsars formed via this route also
receive kicks at its birth and may thus gain energy and be
ejected out from the GC with hyperfast velocities. According to the energy
conservation law, the ejection velocity of pulsar is $v_{\rm ej} \sim
(2\delta v \sqrt{ G \bh /r_{\rm exp}}-\bh G/a_{\rm cap})^{1/2}$, where
$\delta v= |\delta \vec{v}|$ is the kick velocity, $r_{\rm exp}$ is
the distance to the central MBH where the star explodes to form a
pulsar, and $a_{\rm cap}$ is the semimajor axis of the progenitorial
captured star. If $a_{\rm cap} \gg r_{\rm exp}$, we have
\begin{eqnarray}
v_{\rm ej} &\sim & 1500\kms\ \left(\frac{250\AU}{r_{\rm exp}}\right)^{1/4}
                  \times  \nonumber \\
&  & \left(\frac{\bh}{4\times10^6\msun}\right)^{1/4}\left(\frac{\delta
v}{300\kms}\right)^{1/2}.
\end{eqnarray}
For most simulated GC S-stars ($a_{\rm cap} \sim 1000-4000\AU$), they
can turn into hyperfast pulsars and be ejected out to the Galactic
halo/bulge only if they explode at a location close to their
periapsis. 

\end{itemize}

Table~\ref{tab:t1} lists the number of the hyperfast pulsars (with
velocities $\geq 1500\kms$) via the first route and the second route,
respectively. The reference model (Disk-IM2) only produces several to
about $10$ hyperfast pulsars that populate in the Galactic halo, while the
Disk-TH0 model can produce $\sim 50-200$ hyperfast pulsars. Only a
small fraction of these hyperfast pulsars may be beaming toward us, so
that the detectable ones among them are not many. As seen from
Table~\ref{tab:t1}, the majority of the hyperfast pulsars are produced by
the collapse and explosion of massive HVSs; and the contribution from
the collapse and explosion of the captured massive stars is
insignificant as only a small fraction of the massive captured stars
exploded near their periapses can lead to the ejection of hyperfast
pulsars. 

The hyperfast pulsars formed through the above two routes have
velocities substantially larger than that formed in isolation or
binaries. According to equation (\ref{eq:vkick}), the fraction of
pulsars that can be kicked out to hyper-velocities $>1500\kms$ is
about $\sim 10^{-3}$. The current number of known pulsars is $\sim
2000$. Therefore, there could be a few hyperfast pulsars having
velocities $>1500\kms$ formed in isolation or in binaries, and these
hyperfast pulsars could be mixed with those formed through the above
two routes. Future pulsar surveys may find some hyperfast pulsars.
However, it is not easy to identify whether a hyperfast pulsar is
originated from the GC or not as the motion of a hyperfast pulsar
formed through the first route could deviate significantly from the
radial motion because of the supernova kick. 

\section{Conclusions and Discussions}\label{sec:conclusion}

In this paper, we estimate the number and orbital distribution of the
pulsars currently existing in the vicinity of the MBH in the GC, by
assuming that these pulsars are formed from the captured components of
the massive binaries tidally broken up in the vicinity of the MBH.  We
follow the orbital evolution of those pulsars (and its progenitorial
stars) by considering the effects of the limited stellar lifetime, the
supernova kicks, the resonant and the non-resonant relaxations due to
background stars, and the GW radiation. For the model that can
simultaneously re-produce the statistical properties of both the
observed GC S-stars and HVSs (i.e., the Disk-IM2 model), the number of
pulsars with semimajor axis $\leq 4000\AU$ (or $\leq 1000\AU$) is
estimated to be about $97-190$ (or $9-14$). Among them, the closest
one to the central MBH probably has a semimajor axis and pericenter
distance in the range of $116-461\AU$ and $2-228\AU$, respectively. 

The existence of pulsars in the GC is suggested by a number of
indicative observations, including the pulsar wind nebular discovered
by \citet{Wang06}, many X-ray binaries discovered by \citet{Muno05}
and \citet{Bower05} at a distance within one parsec from the MBH, and
a magnetar recently discovered within a parsec distance from the
central MBH \citep[see][]{Rea13, Kennea13, Mori13, Eatough13}, etc.
The discovery of the magnetar is most encouraging because magnetars,
i.e., the magnetized pulsars, are a rare type of pulsars; the
discovery of a single magnetar in the central pc region is almost
exclusively indicating the existence of a substantial number of normal
pulsars within that region.

Recent systematic searches for pulsars did not detect any but gave an
upper limit of $\sim 90$ normal pulsars in orbits within the central
pc of Sgr~A* \citep[e.g.,][]{Macquart10, Eatough12}.  The
non-detection of pulsars in the GC for those searches may be partly
due to that pulsar signals are strongly smeared by the hyper-strong
scattering of radio waves by the turbulent ionized gas within the
central $100\pc$ of Sgr\,A*. The pulse broadening has a strong
frequency dependence, $\propto \nu^{-4}$, making it nearly impossible
to detect any pulsar in the GC at the typical observing frequencies of
$<1.4$GHz. It may be possible to avoid of this smearing problem by
searching pulsars at high frequencies ($>10$GHz).
\citet{Chennamangalam13} revisited the constraint on the number of
potential observable pulsars according to the current observations by
adopting more realistic assumptions; they found a conservative upper
limit $\sim 950$ for the pulsars in orbits within the central pc.

The non-detection of pulsars in the GC is still compatible with our
estimate of the number of the pulsars there ($\sim 97-190$, see
Table~\ref{tab:t1}). According to the method to estimate the number of
observable pulsars developed in \citet[][]{Codres97}, the fraction of
observable pulsars in the GC by the current facilities is $\sim
10^{-2}$ \citep[see Equation~2 in][]{Pfahl04}. As the expected total
number of pulsars in the GC, with semimajor axes $\leq 4000\AU$, is
only $\sim 97-190$, the detectable pulsars would be only $\sim 1-2$
(or even zero) by the current facilities. 

Future radio telescopes, such as the SKA, may be able to detect $10\%$
of the pulsars \citep[see Equation~2 in][]{Pfahl04}. We expect that
ten to twenty (about one) pulsars will be discovered with semimajor
axis $\leq 4000\AU$ ($\leq 1000\AU$) in the GC by the SKA.  Among the
``detected'' pulsars, the closest one to the central MBH probably has
semimajor axis and pericenter distance in the range of $468-1423\AU$
and $21 - 791\AU$, respectively. For other models that can produce
$17$ GC S-stars as observed, the total number of the ``detected''
pulsars, with semimajor axes $\leq 4000\AU$ ( $\leq 1000\AU$), are
about $4-50$ ($1-10$), and the one closest to the central MBH may have
a semimajor axis smaller than that resulting from the Disk-IM2 model.
Long term monitoring of those pulsars close to the central MBH would
be helpful in constraining the environment and the metric of the
central MBH.

Our preferred model also results in about ten hyperfast pulsars with
velocities $\geq 1500\kms$ moving away from the Milky Way, and most of
them are produced by supernovae explosions of the massive HVSs ejected
from the GC. Some of these hyperfast pulsars may be detected by the
SKA in the future. However, it is not easy to confirm that these
hyperfast pulsars are originated from the GC solely by their moving
directions.

\acknowledgements

QY thanks Scott Tremaine for communications on supernova kick. YL
and QY acknowledge the hospitality of the KITP at Santa Babara, where
part of this work was written. This work was supported in part by the
National Natural Science Foundation of China under nos. 11273004 and
11373031, the National Science Foundation under Grant No. NSF
PHY11-25915, the Strategic Priority Research Program ``The Emergence of
Cosmological Structures" of the Chinese Academy of Sciences, Grant No.
XDB09000000, and a grant from the John Templeton Foundation and
National Astronomical Observatories of Chinese Academy of Sciences.
The opinions expressed in this publication are those of the author(s)
do not necessarily reflect the views of the John Templeton Foundation
or National Astronomical Observatories of Chinese Academy of Sciences.
The funds from John Templeton Foundation were awarded in a grant to
The University of Chicago which also managed the program in
conjunction with National Astronomical Observatories, Chinese Academy
of Sciences. 


\end{document}